\documentclass[aps,prb,floats,twocolumn,10pt,floatfix]{revtex4-1}
\usepackage{graphicx}
\usepackage{amssymb}
\usepackage{amsmath}
\usepackage{bbold}

\usepackage{color}
\usepackage[dvipsnames]{xcolor}
\usepackage{slantsc}
\usepackage{float}
\graphicspath{{Figures/}}
\usepackage[caption=false]{subfig}
\usepackage{array}
\usepackage{braket}
\usepackage{scrextend}
\usepackage{hyperref}
\hypersetup{backref,pdfpagemode=FullScreen,colorlinks=true,allcolors=blue}
\usepackage{leftidx}

\begin{document}
	
\title{Machine-learning semi-local density functional theory for many-body lattice models at zero and finite
temperature}
\author{James Nelson}
\email{janelson@tcd.ie}
\author{Rajarshi Tiwari}
\email{tiwarir@tcd.ie}
\author{Stefano Sanvito}
\email{stefano.sanvito@tcd.ie}
\affiliation{School of Physics, AMBER and CRANN Institute, Trinity College, Dublin 2, Ireland}
\date{\today}
	
\begin{abstract}
We introduce a machine-learning density-functional-theory formalism for the spinless Hubbard model in one dimension
at both zero and finite temperature. In the zero-temperature case this establishes a one-to-one relation between 
the site occupation and the total energy, which is then minimised at the ground-state occupation. In contrast, at
finite temperature the same relation is defined between the Helmholtz free energy and the equilibrium site
occupation. Most importantly, both functionals are semi-local, so that they are independent from the size of the 
system under investigation and can be constructed over exact data for small systems. These `exact'
functionals are numerically defined by neural networks. We also define additional neural networks for 
finite-temperature thermodynamical quantities, such as the entropy and heat capacity. These can be either
a functional of the ground-state site occupation or of the finite-temperature equilibrium site occupation. In the first
case their equilibrium value does not correspond to an extremal point of the functional, while it does in the second 
case. Our work gives us access to finite-temperature properties of many-body systems in the thermodynamic limit.
\end{abstract}
	
\maketitle

\section{Introduction}

Machine learning (ML) is a wide field of computer science. It is constructed over a collection of algorithms and 
numerical techniques and aims at recognising and manipulating the patterns hidden in large volumes of 
data.~\cite{Goodfellow-et-al-2016} Although the most impressive progress in the field is related to signal
processing (images, videos, etc.), ML is now becoming a powerful tool in both experimental and computational 
materials science and in engineering.~\cite{butler2018machine,Ramakrishna2019} Examples of applications 
include the design of new molecules,~\cite{Alan2019} the running of self-driving 
labs,~\cite{doi:10.1002/adma.201907801,Steinereaav2211} the construction of ultra-accurate force 
fields,~\cite{doi:10.1002/adma.201902765,Lunghieaaw2210} the prediction of physical properties based
on structure-to-property~\cite{C9CP04489B} or chemistry-to-property~\cite{PhysRevMaterials.3.104405} 
relations, the analysis of data from electron microscopy,~\cite{doi:10.1002/adts.201800037} just to name a few.

In addition to materials science and engineering the use of ML is now widely spread over an impressive range 
of problems in Physics.~\cite{RevModPhys.91.045002} In particular it has been utilised to address the complexity 
of a number of many-body models. For these an exact solution typically does not exist, except for a few limiting cases, 
while exact diagonalisation is bound by the severe scaling of the Hilbert space with the system size, and hence by 
the computational costs. For instance, ML has been used to find ground-state properties and observables of 
fermion-boson coupled Hamiltonians,~\cite{saito2017solving} disordered Hubbard-Anderson models~\cite{ma2019machine} 
and a variety of spin models.~\cite{Nieva2019,PhysRevB.99.024423,PhysRevE.97.032119} At the
same time some progress has been made in classifying quantum states at finite temperature and thus identifying 
phase transitions.~\cite{Carrasquilla2017,PhysRevE.97.013306}. Finally, restricted Boltzmann machines have been
constructed to represent and propagate in time wavefunctions,~\cite{carleo2017solving} a work that has stimulated 
the introduction of open-source software for the study of many-body models.~\cite{NetKet} 

Several numerical methods have been designed to solve many-body problems for systems, whose size makes 
them not accessible by exact diagonalisation. If the interest lies in ground-state properties, density matrix 
renormalisation group~\cite{white1992density} and variational Monte Carlo~\cite{foulkes2001quantum} methods are 
valuable options, while at finite temperature continuous-time Monte Carlo~\cite{gull2011continuous} and dynamic 
mean field theory~\cite{kotliar2006electronic} can be considered. These methods, depending on the nature of their 
implementation/design, usually suffer from (i) restrictions to low or high dimensions, (ii) the fermionic sign problem, 
(iii) poor reliability at either low or high temperatures. In any case, although some schemes may allow one to investigate 
relatively large systems, their numerical overheads are still significant. This means that information concerning the 
thermodynamic limit of the various models and about the interplay between interaction and disorder remain difficult to 
access.

In this context functional-based methods, such as density functional theory (DFT), deserve a class on their own.
These are based on the two Hohenberg-Kohn theorems~\cite{hohenberg1964inhomogeneous}, which have also 
been extended to lattice models.~\cite{PhysRevLett.56.1968,PhysRevB.52.2504,Coe_2015} The core idea is that
the total energy can be expressed as a universal functional of the single-particle electron density, or its lattice
equivalent, and that such functional has a minimum at the ground-state density, where it gives the ground-state 
energy. For interacting theories the universal functional is not known, although excellent approximations exist and
a number of exact constrains have been rigorously proved. Most importantly for the discussion here, the functional formulation 
of the many-body problem appears ideal for a ML approach. In fact, one can construct a dataset of either exact or 
approximated results and try to learn the one-to-one relation between the electron density and the energy. 
Several examples of this approach exist for the actual Coulomb-interaction DFT~\cite{doi:10.1021/acs.accounts.0c00742} for both the 
non-interacting~\cite{PhysRevLett.108.253002,doi:10.1002/qua.25040} and 
interacting~\cite{PhysRevB.94.245129,Schmidt2019} case in one dimension, and in two dimensions for 
selected external potentials.~\cite{PhysRevA.96.042113,PhysRevA.100.022512} There are also attempts in three
dimensions, in particular at constructing maps between the potential and the electron density~\cite{Burke2017} and 
between the electron density and the total energy,~\cite{Yao2016,Burke2020} and in learning the self-consistent electron
density in Kohn-Sham DFT.~\cite{Zepada2019}

In our previous work we have constructed an exact machine-learning DFT functional for the one-dimension Hubbard 
model using neural networks,~\cite{nelson2019machine} a strategy also used by Moreno et al. to reconstruct
the ground-state wave-function.~\cite{moreno2020deep} We have then demonstrated that such functional satisfies 
both the Hohenberg-Kohn theorems.~\cite{hohenberg1964inhomogeneous} The network, trained over exact-diagonalisation
results for systems defined by a random single-particle potential, establishes a one-to-one relation between the sites 
occupation and the ground-state total energy. Since the site occupation of every site (the total 
charge density) is used to define the functional, one has to construct a new network, namely a new functional, for 
every different system size and every different filling factor. This means that, although the functional is exact to numerical 
precision it is of little practical use. 

Here we circumvent the problem by constructing a semi-local functional, which requires only local knowledge of the 
site occupation. This can now be trained over small systems and be used to predict the ground-state properties of 
systems of any size. The same formalism is then extended to finite-temperature, a task that is approached in two different 
ways. Firstly, by using one of the results of the Hohenberg-Kohn theorems, we construct neural networks that relate
the ground-state density to the thermal average of a number of operators. These define the finite-temperature 
equilibrium site occupation, energy, entropy and specific heat. In this case, although such quantities are functional
of the ground-state site occupation, they are not found at the functional minimum, namely they cannot be determined 
by variational principle. In contrast, the second approach consists in extending DFT to the canonical ensemble as
proposed a long time ago by Mermin.~\cite{PhysRev.137.A1441} The one-to-one correspondence is now
between the finite-temperature equilibrium site occupation and the Helmholtz free energy. Also in this case
we construct a universal semi-local functional, which is minimised at the equilibrium site occupation.

The paper is organised as follows. In the next section we introduce our ML models, by presenting the semi-local 
functional and the representation used for the site occupation, by introducing functionals for finite-temperature 
thermodynamic properties and by formulating the extension of DFT to the canonical ensemble. In this section we
will also discuss the details of the neural networks constructed. Then we present our results
focusing first on the zero-temperature limit and then to its finite-temperature extension. In particular we discuss
the actual level of non-locality required by the functionals and analyse the homogeneous limit. Finally, we show how
the functional can be minimised to find the ground-state density (or the finite temperature equilibrium density), and 
how this procedure allows us to extract finite-temperature properties of the model for systems of any size. 
Finally, we conclude.

\section{Construction of the models}

\subsection{Zero-temperature lattice DFT: the semi-local density approximation}
Our analysis applies to any lattice model with Hamiltonian of the form,
\begin{equation}\label{H}
\hat{H} = \hat{H}_0 + \hat{H}_\mathrm{kin} + \hat{H}_\mathrm{int}\:,
\end{equation}
where $\hat{H}_0$ is a single-particle potential, $\hat{H}_\mathrm{kin}$ is the kinetic energy and $\hat{H}_\mathrm{int}$ 
is some form of many-body interaction. In particular here we focus our attention on the spinless non-local 
Hubbard model in one dimension. This is implemented with periodic boundary conditions, namely for finite rings 
comprising $L$ sites. The corresponding Hamiltonian, $\hat{H}_V$, thus reads,
\begin{equation}\label{H_spinless}
\hat{H}_V = \sum_{i=1}^{L}\epsilon_i\hat{n}_i
-t\sum_{i=1}^{L}(\hat{c}_i^\dagger\hat{c}_{i+1}+ \hat{c}_{i+1}^\dagger\hat{c}_{i}) + 
V\sum_{i=1}^L\hat{n}_{i+1}\hat{n}_{i} \:,
\end{equation}
where $\hat{c}_i^\dagger$ ($\hat{c}_i$) is the creation (annihilation) operator for a spinless electron at site 
$i$, $\hat{n}_{i}=\hat{c}_i^\dagger\hat{c}_i$ is the number operator, $\epsilon_i$ are the on-site energies,
$V$ is the non-local Hubbard parameter and $t=1$ is the hopping integral that sets the energy scale of
the problem. The results presented here are for the half-filling case, namely the total number of electrons
is $N_{e}=L/2$.

Since we are going to compare quantities for lattices of different size, it is convenient to define any observable
in term of its density, namely by dividing it by the number of lattice sites. For instance, the energy density is
simply, $e\equiv E/L$, where $E$ is the total energy. It was demonstrated some time ago that the Hohenberg-Kohn 
theorem~\cite{hohenberg1964inhomogeneous} can be extended to lattice 
models,~\cite{PhysRevLett.56.1968,PhysRevB.52.2504,Coe_2015} once the appropriate single-particle
density is defined. In the case of the spinless Hubbard model the fundamental quantity is the sites occupation,
$\{n_i\}$, defined as the expectation value of the number operator, $n_i=\langle\hat{n}_i\rangle$, over all the 
sites. As a matter of notation here we represent the ground-state expectation value of the generic operator 
$\hat{O}$ as $\langle\Psi_0|\hat{O}|\Psi_0\rangle=\langle\hat{O}\rangle_0$, where $\Psi_0$ is the many-body 
ground-state wave-function. The energy density functional (in this case an actual function) can then be written as,
\begin{equation}
e[\{n_i\}] = f_V[\{n_i\}] + \frac{1}{L}\sum_{i=1}^{L}\epsilon_in_i\:,
\end{equation}
where $f_V[\{n_i\}]$ is a universal functional of $\{n_i\}$ and it is defined for every value of $V$ (there is a
functional for every value of $V$). In our previous work we have shown that a numerically exact functional 
could be learnt by using ML.~\cite{nelson2019machine} This satisfies both the two Hohenberg-Kohn theorems, 
namely for a given ground-state density it yields the external potential, and it is variational in the site occupation. 
The second statement implies that the minimum of the functional is found at the ground-state site occupation,
$n_i^0=\langle\hat{n}_i\rangle_0$, where it returns the ground-state energy density, $e^0=e(\{n_i^0\})$.
Such ML functional depends on the sites occupation at every sites, therefore it is strictly defined for a particular 
system size and electron filling factor. Thus, the functional is of little practical use, since it needs to be constructed 
for any specific system under investigation, an operation that is limited by the computational ability of solving 
exactly the full many-body problem. 

\begin{figure}
	\centering
	\includegraphics[width=\columnwidth]{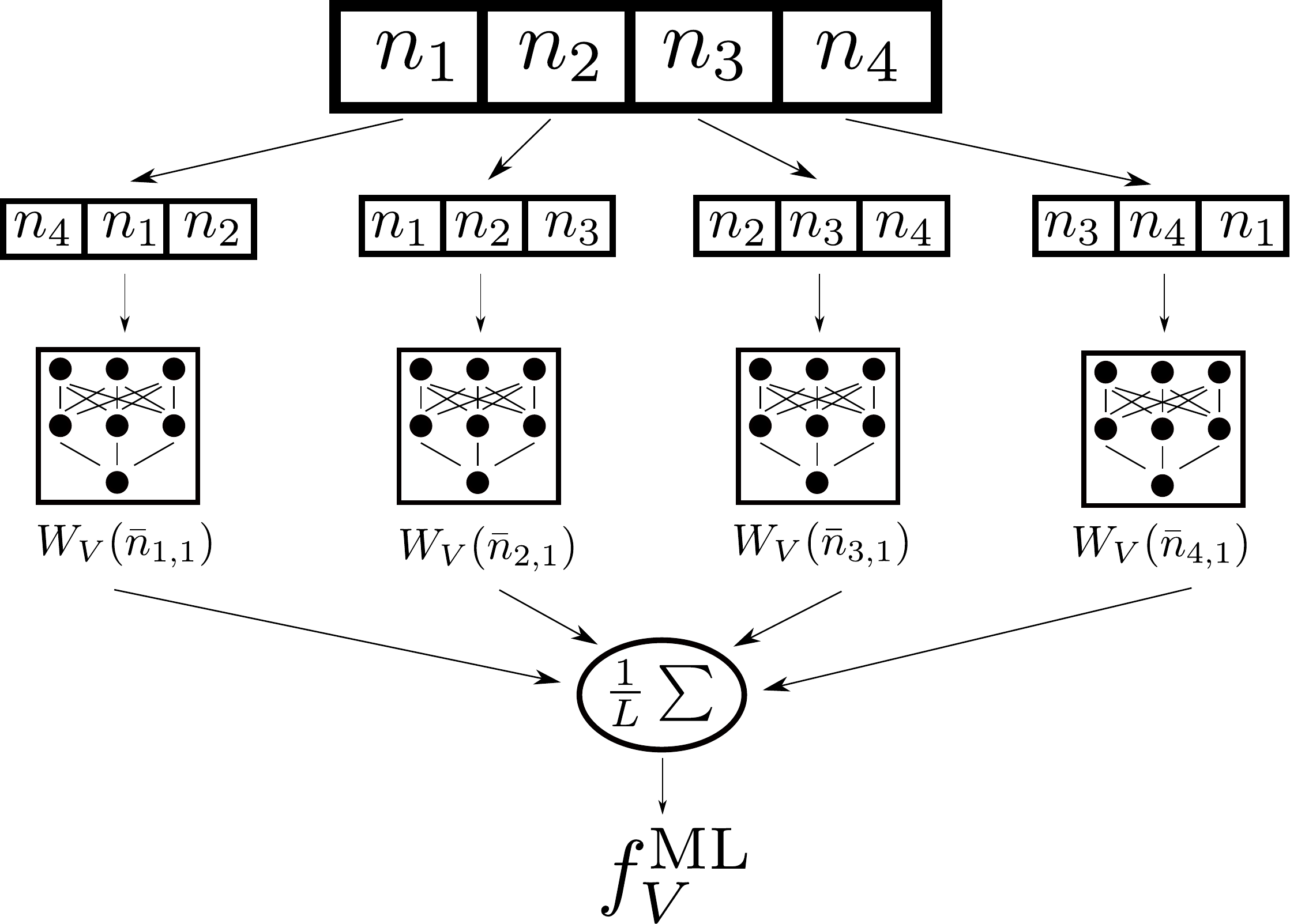}
	\caption{An illustration of how the semi-local functional, $f_V^\text{ML}[\{n_i\}]$, is constructed for a 4-site 
	ring with $a=1$. In this case four occupations, $\{n_1,n_2,n_3,n_4\}$, define uniquely the total energy. 
	This is written as the site average average of four $W_V(\bar{n}_{i,a})$ contributions, constructed via a ML 
	neural network, each one of them depending only on three site occupations.}
	\label{Fig1}
\end{figure}

Here we overcome such drawback by introducing a new semi-local functional that is, by construction, 
lattice-size independent. This is defined as
\begin{equation}
f_V^\text{ML}[\{n_i\}] = \frac{1}{L}\sum_{i=1}^{L}W_V(\bar{n}_{i,a})\:,
\end{equation}
where $W_V(\bar{n}_{i,a})$ is the energy associated to the $i$-th site. In turn, $W_V(\bar{n}_{i,a})$ depends 
on the `local' site occupation,
\begin{equation}
\bar{n}_{i,a} = \{n_{i-a}, n_{i-a+1}, ... , n_i , ... , n_{i+a-1}, n_{i+a}\}\:,
\end{equation}
meaning that the energy associated to site $i$ depends on the occupation at site $i$ and on that at the
first $a$ sites around it (overall, it depends on the site occupation at $2a+1$ sites). Thus $a$ defines the locality 
of the functional, $f_V^\text{ML}[\{n_i\}]$, which is simply the average of $W_V(\bar{n}_{i,a})$ over all the 
sites. Fig.~\ref{Fig1} illustrates an example of how to construct the functional for a ring of 4 sites and a local 
density of range $a=1$.

Recently, a conceptually similar way to construct ML models for extensive quantities has been brought forward by 
Mills and co-workers,~\cite{mills2019extensive} who represented the physical space of the model (e.g. the atomic sites)
across a set of non-overlapping regions. A common neural network defined for such region is then averaged over 
the entire space to yield the desired extensive quantity. Our approach is similar in spirit, with the main difference
being the use of a site-centered representation of the local quantity of interest (the site occupation). 
As well as being lattice-size independent this approximation has site symmetry automatically built into it, 
namely it will predict, by construction, the same energy for densities formed by translating the onsite potential, 
$\{\epsilon_i\}\rightarrow \{\epsilon_{i+1}\}$.

\subsection{Zero-temperature lattice DFT: thermodynamic quantities}

In lattice DFT (Coulomb DFT) the Hohenberg-Kohn theorem establishes a one-to-one correspondence between 
the ground-state site occupation (the single-particle electron density) and the on-site energies (the external potential). 
It follows that $\{n_i^0\}$ completely determines the Hamiltonian of the system, hence its solution. Thus, one can conclude 
that the entire many-body excitation spectrum is a functional (unknown) of the ground-state site occupation, namely
\begin{equation}\label{Enfunc}
|\Psi_m\rangle=|\Psi_m[\{n_i^0\}]\rangle, \:\:\:\:\:E_m=E_m[\{n_i^0\}]\:,\:\:\:\:\:\:\:\:\:\:{\mathrm{for\:all}\:m}\:,
\end{equation}
where $E_m$ is the energy of the $m$-th many-body state and $|\Psi_m\rangle$ the associated wave-function.

Now take a generic operator, $\hat{O}$, and calculate its thermal average, $\langle\hat{O}\rangle_T$, in the canonical 
ensemble. This reads
\begin{equation}\label{OT}
\langle\hat{O}\rangle_T=\sum_n \langle\Psi_n|\hat{O}|\Psi_n\rangle\:\frac{\mathrm{e}^{-\beta E_n}}{Z}\:,
\end{equation}
where $Z=\sum_n\mathrm{e}^{-\beta E_n}$ is the partition function, and $\beta=1/k_\mathrm{B}T$ with 
$k_\mathrm{B}$ being the Boltzmann constant and $T$ the temperature. It then follows, from Eq.~(\ref{Enfunc}), 
that also $\langle\hat{O}\rangle_T$ is a functional of the ground-state density, $\langle\hat{O}\rangle_T[\{n_i^0\}]$. 
Clearly in this case the variational principle is not established, meaning that $\langle\hat{O}\rangle_T$ does
not necessarily have an extremal point at $\{n_i^0\}$. However, one can still define a map, $f$, 
between ground-state site occupation and temperature, and $\langle\hat{O}\rangle_T$,
\begin{equation}\label{maps}
f: [\{n_i^0\},T] \longrightarrow \langle\hat{O}\rangle_T\:.
\end{equation}

We have thus constructed a number of neural networks to numerically estimate such mapping. These are also
built within the semi-local approximation, namely they have the form,
\begin{equation}\label{MLT}
\langle \hat{o}\rangle^\text{ML}_T = \frac{1}{L}\sum_{i=1}^{L}W_V^O(\bar{n}_{i,a}, T)\:,
\end{equation}  
where $\hat{o}=\hat{O}/L$ is the operator density and where now the neural network, $W_V^O(\bar{n}_{i,a}, T)$,
depends also on the temperature. In particular we have considered the entropy density, $\hat{s}$, the energy
density, $\hat{e}$, and the heat capacity density, $\hat{c}$, which read respectively
\begin{equation}
\langle \hat{s}\rangle_T=-\frac{1}{L}\sum_n p_n\ln p_n 
=\frac{1}{L}\frac{\partial}{\partial T}(k_\mathrm{B}T\ln Z)\:,
\end{equation}
\begin{equation}
\langle \hat{e}\rangle_T=\frac{1}{L}\sum_n \langle\Psi_n|\hat{H}|\Psi_n\rangle\:\frac{\mathrm{e}^{-\beta E_n}}{Z}
=-\frac{1}{L}\frac{\partial \ln Z}{\partial\beta}\:,
\end{equation}
\begin{equation}
\langle \hat{c}\rangle_T=\frac{\partial \langle e\rangle_T}{\partial T} 
=\frac{1}{L}\frac{1}{k_\mathrm{B}T^2}\frac{\partial^2 \ln Z}{\partial\beta^2}\:,
\end{equation}
where $p_n=\mathrm{e}^{-\beta E_n}/Z$.

Note that in general the ground-state site occupations determine the energy only up to a constant, $E_\alpha$. 
A constant shift in the many-body energy eigenvalues, $E_n\rightarrow E_n+E_\alpha$, transforms the partition 
function as $Z \rightarrow Z\exp{(-\beta E_\alpha)}$, but leaves invariant both $\langle \hat{s}\rangle_T$ and 
$\langle \hat{c}\rangle_T$. The same is not true, however, for $\langle e\rangle_0$ and $\langle e\rangle_T$, so 
that one has to constrain the on-site energies to a constant. In our previous work\cite{nelson2019machine} such 
constraint was imposed by setting one of the on-site energies to zero. Here, in order to preserve the symmetry 
of the single particle potential, we impose that the on-site energies have always zero mean. We implement such 
condition by shifting the energy eigenstates by $E_\alpha=-(N_e/L)\sum_j\epsilon_j$, since this transforms the 
total energy as $E \rightarrow E + N_e E_\alpha$. 

\subsection{Finite-temperature lattice DFT}

In conventional DFT the Hohenberg-Kohn theorems can be generalised to both the canonical and the grand 
canonical ensemble.\cite{PhysRev.137.A1441} Their translation to lattice DFT for the specific case of the
spinless Hubbard model and the canonical ensemble reads as follows. One can define a universal functional, 
$g_V[\{n_i\}]$, independent from the one-site energies, $\{\epsilon_i\}$, such that 
\begin{equation}
h(\{n_i\}) = g_V[\{n_i\}] + \frac{1}{L}\sum_{i=1}^{L}\epsilon_in_i\:
\end{equation}
is minimum and equal to the Helmholtz free energy density, $h$, associated to $\{\epsilon_i\}$, when $\{n_i\}$ is 
the equilibrium site occupation in the presence of $\{\epsilon_i\}$. This essentially means that at the minimum
of the functional one has
\begin{equation}
h[\{n_i^T\}] = \langle \hat{e}\rangle_T-T\langle \hat{s}\rangle_T\:,
\end{equation}
with the equilibrium site occupation defined as [see Eq.~(\ref{OT})],
\begin{equation}
n_i^T=\langle \hat{n}_i\rangle_T=\sum_n \langle\Psi_n|\hat{n}_i|\Psi_n\rangle\:\frac{\mathrm{e}^{-\beta E_n}}{Z}\:.
\end{equation}

Also in this finite-temperature extension of DFT the universal functional can be constructed by using
a neural network and a semi-local approximation, namely we can define
\begin{equation}
g_V^\text{ML}[\{n_i\}] = \frac{1}{L}\sum_{i=1}^{L}X_V(\bar{n}_{i,a})\:,
\end{equation}
which again has a variational minimum at $\{n_i^T\}$.

Finally, note that the equilibrium site occupation at site $i$ can be simply viewed as the thermal
average of the number operator, $\hat{n}_i$. Hence, we can construct a machine-learning model
that, given $n_i^0$ and $T$, returns $n_i^T$ [see Eq.(\ref{MLT})], namely
\begin{equation}\label{nedensity}
n_i^T=\langle \hat{n}_i\rangle^\text{ML}_T = W_V^{n_i}(\bar{n}_{i,a}, T)\:.
\end{equation}  
Since all the functionals constructed here are semi-local, we have now a tool to approach the study
of thermodynamic quantities for a disordered many-body system in the limit $L\rightarrow\infty$. In 
fact we have two options. On the one hand, one can considered finite-temperature lattice DFT and
compute, by variational principle, both $\{n_i^T\}$ and $h(\{n_i^T\})$. On the other hand, we can 
use zero-temperature DFT to determine $\{n_i^0\}$ and then, by employing the machine-learning 
maps of Eq.~(\ref{maps}), all the thermal averaged observables. Indeed, one can also construct
a `hybrid' approach where $\{n_i^T\}$ is derived from $\{n_i^0\}$ via Eq.~(\ref{nedensity}) and then
used to determine the Helmholtz free energy.

\section{Results}
In order to test the accuracy of our semi-local functionals we have generated datasets for $L=6, 10, 14$ and 
$V=1, 2, 4$, by exact diagonalization. The training and validation sets, used respectively to train the model 
and to enforce early stopping, contain data where the on-site energies are drawn from several uniform 
distributions in the interval, $[-W,W]$, with $W$ ranging from $2$ to $8$. In total for any given $L$ and $V$ 
there are 24,000 random on-site energy realizations in the training set and $8,000$ in the validation one. Finally, 
the test set contains 500 configurations generated for $W=4$. 

\subsection{Construction of the semi-local functional at zero temperature}
\begin{figure}[t]
	\centering
	\includegraphics[width=\columnwidth]{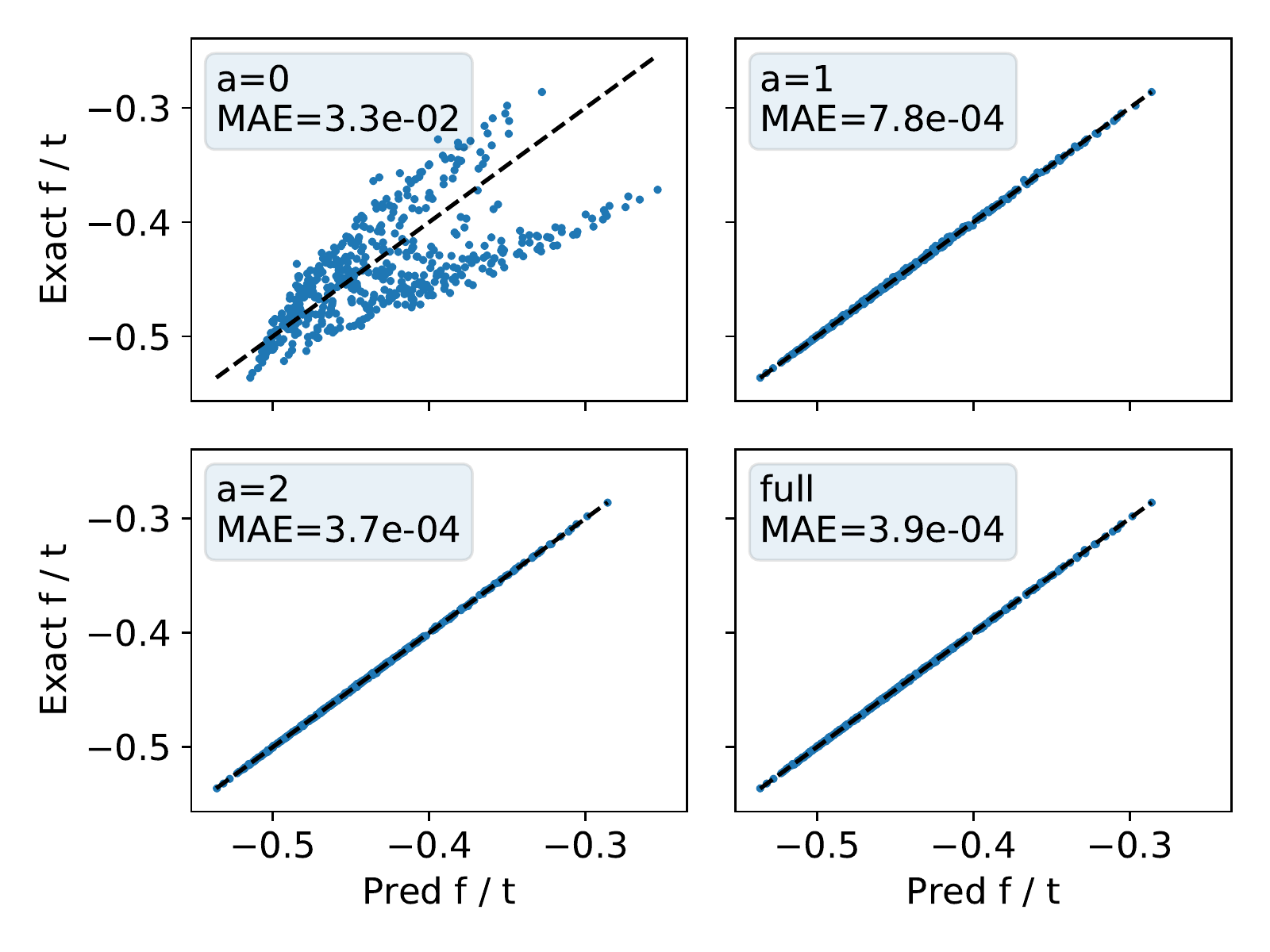}
	\caption{(color on-line) Accuracy of the semi-local functional for ground-state DFT as a function of the locality parameter, $a$. Here we plot 
	the value of the exact universal functional (computed from exact diagonalization) against the prediction obtained with our ML semi-local formulation.
	Results are for $L=6$ and $V=1$. The box inside each plot reports the $a$ value and the MAE, with `full' corresponding to 
	a functional, where the entire site occupation vector is used (completely non-local case). Data are provided here for the test set and each
	graph contains 500 points.}
	\label{fig:fig2_local}
\end{figure}

In order to learn the zero-temperature universal functional, $f_V$, we have used a fully connected neural network, 
constructed by using PyTorch.~\cite{paszke2019pytorch} Weight sharing is implemented to ensure that the $L$ available 
local energies, $W_V\{\bar{n}_{(i,a)}\}$, are the same, and the `Adam' optimizer is chosen to update the models. The first
validation of our strategy consists in investigating the accuracy of a functional constructed for a system of $L$ sites against
new on-site energy configurations for systems with the same number of sites. Namely, we first check how the ML functional
trained for $L$ performs against unknown $L$-site systems. This must be done as a function of the locality parameter, $a$, so that 
the efficacy of the semi-local approximation is established. Our results for $V=1$ and $L=6$ are shown in Fig.~\ref{fig:fig2_local}, 
where we plot the functional computed from exact diagonalization against our semi-local ML estimation for a collection of
different systems. The figure of merit here is the mean absolute error (MAE) and similar trends have been observed for 
different values of $V$ and $L$. 

From the figure it is clear that, although a completely local approximation, $a=0$, is insufficient to approximate the functional,
already a moderate degree on non-locality, $a=1$, returns us a MAE just below 10$^{-3}$, corresponding to an average error of 
the order of 0.2\%. Extending further the range of the functional improves the description, to a point that the $a=2$ case presents 
an error, $\sim$0.1\%, numerically indistinguishable from that of the fully non-local ML functional. 
\begin{figure}
	\centering
	\includegraphics[width=\columnwidth]{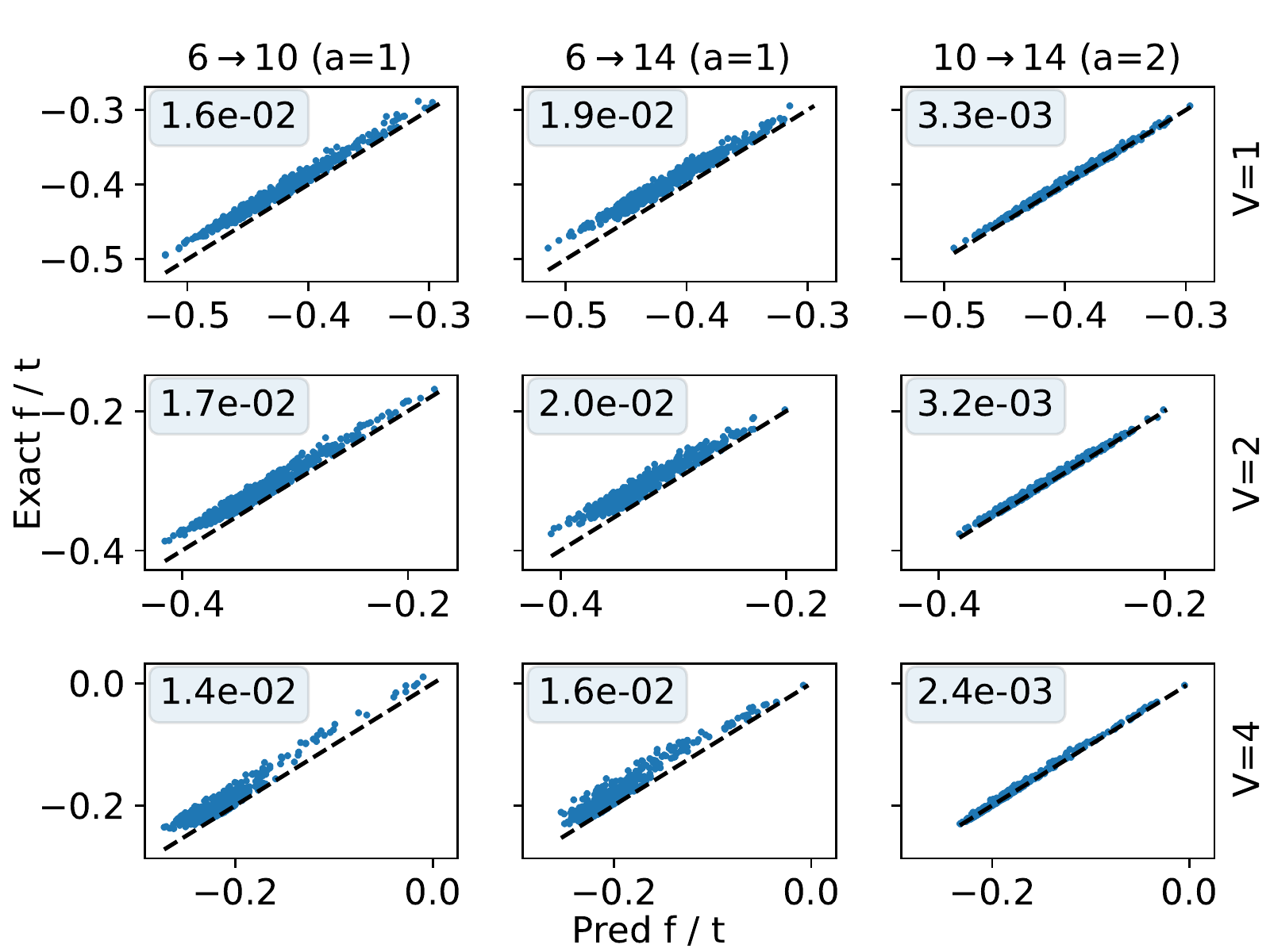}
	\caption{(color on-line) Accuracy of the semi-local ground-state DFT functional trained for small systems and tested for larger ones as 
	a function of the locality parameter, $a$. Here we plot the value of the exact universal functional (computed from exact diagonalization) 
	against the prediction obtained with our ML semi-local formulation. In the first column the functionals have been trained on rings of $L=6$ 
	sites and are used to predict $L=10$ data. The second column is for functionals trained on 6-site rings and predicting $L=14$ data. 
	Finally, in the third column the functionals have been trained on 10-site rings and used to predict $L=14$ data. The rows correspond 
	to different values of $V$. The value of $a$ is the same for each column.}
	\label{fig:fig3_gen}
\end{figure}

Since the local representation is lattice-size independent and the universal functional is semi-local, in principle one can use small, cheaply 
generated, systems as training set, and then predict quantities for larger systems. The results of this exercise are presented in Fig.~\ref{fig:fig3_gen},
which compares the accuracy of ML functionals trained over rings of $L$ sites at predicting, $f_V$, for rings of $L^\prime$ sites (symbolically 
$L\rightarrow L^\prime$), with $L^\prime>L$ and for various values of $V$ and $a$. We can notice that when the ML training takes place over 
relatively small rings ($L=6$ in this case) the ML estimate of $f_V$ appears to be larger in absolute value than the exact result, namely all the 
points lie above the perfect-agreement line. This is, for instance, the case of training on a $L=6$ lattice and predicting for $L=10, 14$. 
Such overestimation is systematic and independent from the interaction strength $V$. However, the error is immediately corrected when training 
the network on larger lattices and in fact the $10\rightarrow14$ case already offers an error below 1\%, with a non-locality of $a=2$, regardless of
$V$.

The inability to train the model over very small rings is rooted into our representation, which does not describe the homogeneous case. Consider,
in fact, the situation where $\epsilon_i=\epsilon_0$, namely all the on-site energies are identical. Translational invariance then imposes that 
$n_i=N_e/L$ for any site $i$, so that $\bar{n}_{i,a} = \bar{n}_{a} = \{N_e/L, ..., N_e/L\}$. The ML functional then has the following simple form,
\begin{equation}
	f_V^\text{ML}[\{n_i\}] = \frac{1}{L}\sum_{i=1}^{L}W_V(\bar{n}_{i,a}) = W_V(\bar{n}_{a})\:,
\end{equation}
namely it returns the same energy density regardless of the system size, $e(\{n_i\})=W_V(\bar{n}_{a})+\epsilon_0N_e/L$. This is clearly incorrect,
even for the non-interacting case ($V=0$). Thus, our semi-local functional formulation is unable to describe the kinetic energy of rings of arbitrary
size. Such error becomes progressively smaller as one trains the network on large rings, since the difference in kinetic energy density between a 
ring with $L$ sites and one with $L+n$ reduces with $L$. Eventually the error is completely eliminated in the thermodynamic limit, where the kinetic
energy density becomes $4t/\pi\sin\left(\frac{N_e\pi}{2L}\right)$.

One can then design a number of alternative strategies in order to eliminate or mitigate the error made on the non-interacting kinetic energy density.
For instance we can explicitly subtract from the universal functional the non-interacting energy, a single-particle quantity easy to compute. This is 
effectively what usually done in conventional DFT, when defining the exchange and correlation functional. Here, we have decided not to include any 
correction, but simply train over larger rings (e.g. $L=10$), for which our numerical analysis demonstrates that the error is relatively small (see 
Fig.~\ref{fig:fig3_gen}). Most importantly, it needs to be noted that the kinetic contribution to the total energy gets smaller as disorder is included 
(as we depart from the homogeneous limit) and the interaction gets larger. 

\begin{figure}
	\centering
	\includegraphics[width=\columnwidth]{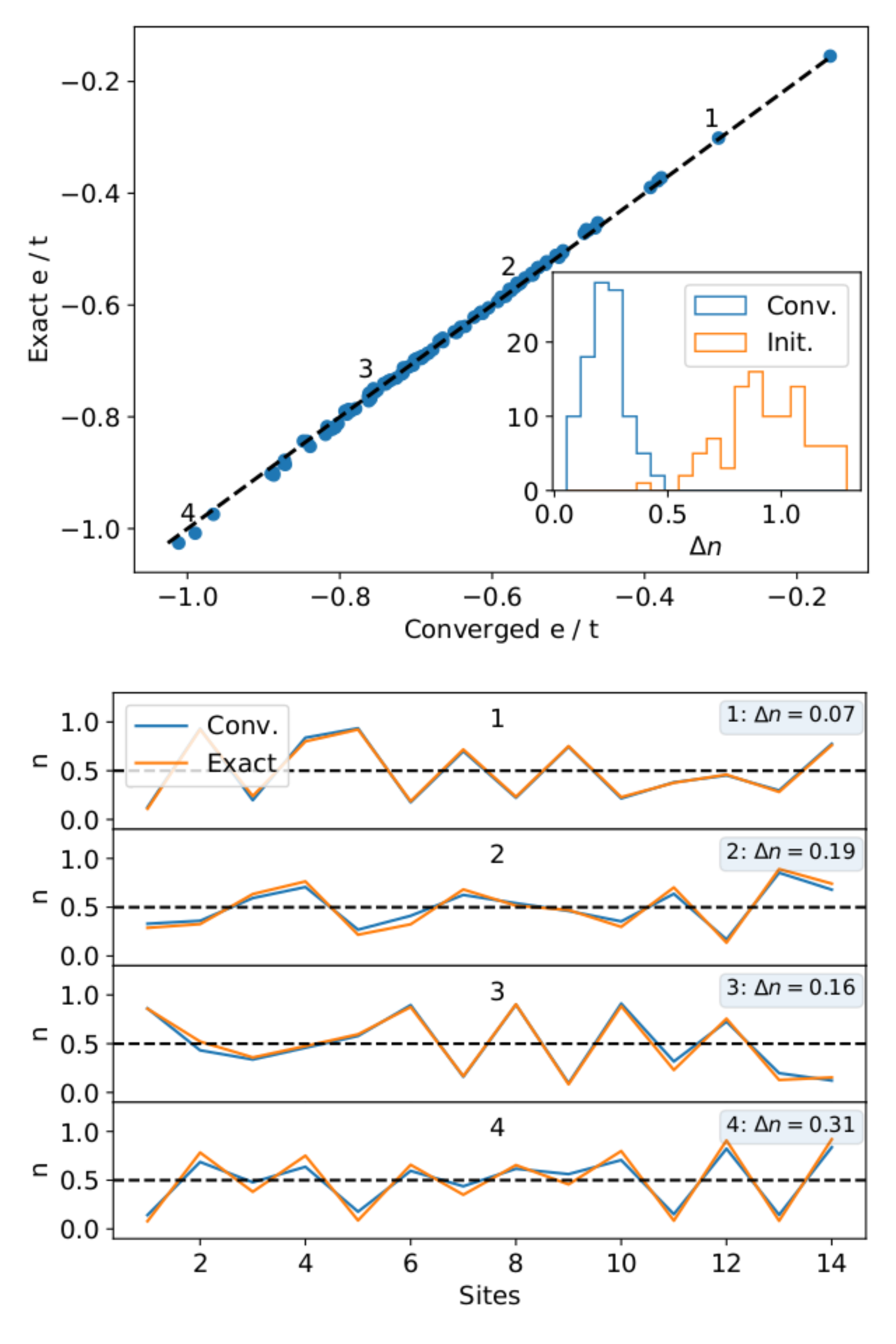}
	\caption{(color on-line) Demonstration of the variational principle for the ML functional. In the top panel we show the energy density computed
	by minimization with respect to the site occupation. Our network has been trained for $L=10$ and used for a ring containing 14 sites ($V=1$). 
	The figure contains 100 different disorder realizations with $W=4$. The insert shows histograms of the euclidean distances $\Delta n$, of the 
	initial and converged site occupation from the exact one. The lower panel presents a selection of the converged occupations compared against 
	the exact value. These are taken for different ground-state energy densities as indicated by the numbers in the upper panel. The overall MAE
	of the site occupation of 0.0047.}
	\label{fig:fig4_gd}
\end{figure}

The tests presented so far only concern the energy, namely we have computed the value of $f_V$ at the exact charge density obtained
with exact diagonalization. This simply demonstrates that our neural network is a good approximator for $f_V$, but does not mean that we
have a variational theory. What we need to show instead is that a network trained on a small lattice can be minimised by the ground state
charge density also for larger lattices, and that at the minimum the energy density returns its exact ground-state value. This essentially means that the 
functional is variational and it is transferrable from small to large rings (namely, the semi-local approximation works). Figure~\ref{fig:fig4_gd} 
provides such demonstration. By taking the $V=1$ case, we train a neural network on a $L=10$ ring. Then, by starting from the homogeneous 
site occupation distribution, $n_i=N_e/L$, we minimize the energy density for a ring of 14 sites. The minimization is performed by using gradient descent 
with momentum \cite{Goodfellow-et-al-2016} (learning rate, $0.002$, and momentum, $0.9$), and then the converged site occupations are 
used to compute the energy density. The top panel of Fig.~\ref{fig:fig4_gd} compares the ML-computed energy density with the exact diagonalization result
for 100 random rings. In the inset we show an histogram of the euclidean distances between the initial and the converged site occupations, and the 
exact $\{n_i^0\}$. The agreement appears quite good, with a MAE of 0.0047, and a generally accurate estimate for the ground-state
occupation. Some examples of converged site occupations are also included in the figure. 

\subsection{Construction of the semi-local finite temperature maps}
In this section we establish one-to-one mappings between the ground-state occupation and a number of thermodynamic quantities. For this
task we have used two fully connected neural networks; the first predicts the three scalar thermodynamic functions, $\langle \hat{s} \rangle_T$, 
$\langle \hat{e} \rangle_T$ and $\langle \hat{c} \rangle_T$, and the second, the equilibrium site occupation $\{\langle n_i \rangle_T\}$ (a vector). 
For all of the thermodynamic predictions, our validation and test sets consist of the function to predict for a particular system at a particular temperature. 
In contrast, the training set contains data for several temperatures for the same system, so to increase the size of the set. The temperatures is randomly 
chosen between $1$ and $2$, with the Boltzmann constant set to one (natural units). 
\begin{figure}
	\centering
	\includegraphics[width=0.98\columnwidth]{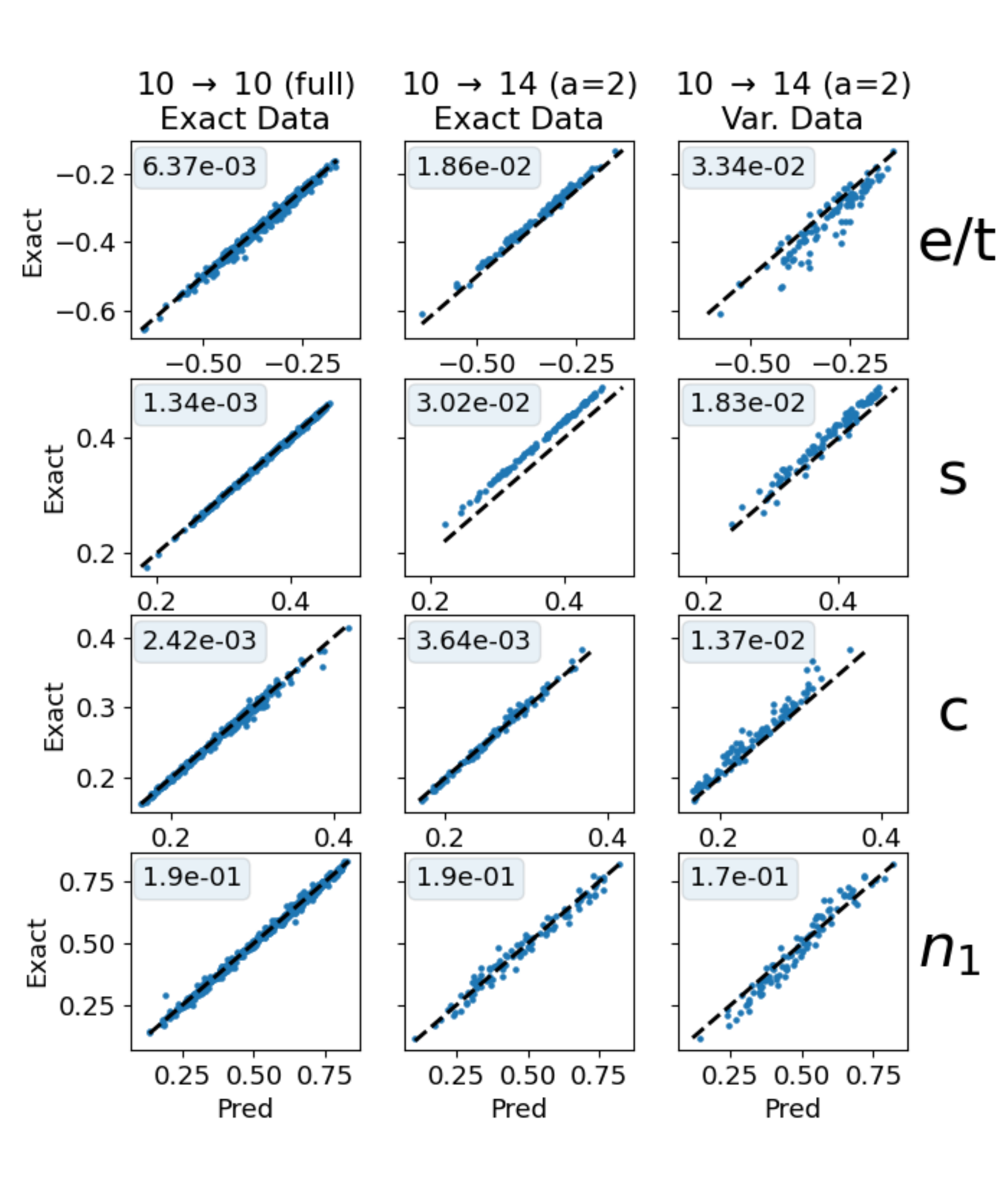}
	\caption{(color on-line) Demonstration of the one-to-one mapping between ground-state charge density and thermodynamic quantities:
	$\langle \hat{e} \rangle_T$, $\langle \hat{s} \rangle_T$, $\langle \hat{c} \rangle_T$ and $\{\langle n_i \rangle_T\}$. For $\{\langle n_i \rangle_T\}$
	we plot only site $i=1$. Approximated values are plotted against exact results for systems at random temperatures in the range $T\in[1,2]$ 
	obtained with $V=1$ and $W=4$. The left-hand side column is for $L=10$ rings and fully non-local models trained on rings of same size. 
	In this case the site occupation is taken from exact diagonalization. The middle column is for semi-local models with $a=2$, where the 
	networks are trained on $L=10$ rings and predict for $L=14$ ones. In this case the ground-state charge density used is the exact one. 
	The right-hand side column depicts results for the same model, but where now the charge density is obtained by variational principle from the ground-state 
	functional. In each panel we report the MAE.}
	\label{fig:fig5_genFT}
\end{figure}

Figure~\ref{fig:fig5_genFT} shows the results obtained by using this method. In particular we present a comparison between the ML-estimated
and the exact quantities for a number of random samples and temperatures, when three different computational strategies are adopted. Data in the 
left-hand side column are for $L=10$ rings obtained with neural networks trained on rings of the same size. In that case the entire site-occupation
vector is used as input, namely the model is completely non-local. The ground-state site occupation used in the ML model is taken from exact
diagonalization, namely it is exact.  Thus, this first case aims at understanding whether ML can construct the one-to-one correspondence between 
ground-state occupation and thermodynamical quantities in the canonical ensemble. In contrast, the middle column shows results for networks trained 
on $L=10$ rings and predicting the thermodynamics quantities for $L=14$ rings. The locality parameter is chosen to be $a=2$, but the site occupation
used to evaluate the various quantity is against the exact one. Thus, this set effectively tests the accuracy of the one-to-one mappings. Finally, the
left-hand side column is for the same models but now evaluated at the site occupation obtained by minimizing the total energy density with the $f_V$ 
built for the $L=10$ case. Such last case, then explores whether or not a semi-local approximation to the thermodynamics exists and whether this is 
accessible from the site occupation obtained by variational principle.

From the figure it clearly emerges that ML is indeed able to predict the mapping, and this is significantly more accurate when the ground-state
site occupation is exact. A semi-local approximation for such mapping appears to be possible, although the error is certainly larger. In particular,
it is clear that there are two main sources or error. The first one is related to the semi-local approximation, so that the thermodynamical quantities
computed at semi-local level are less accurate than those for the completely non-local model, even when these are evaluated at the exact site
occupation (compare the left-hand side column with the middle one). Then there is additional error originating from the use of an approximated 
approximated site-occupation (compare the middle column with the right-hand side one). Notably, the entropy seems to be the quantity suffering 
the most from the absence of the correct homogeneous limit in the semi-local models, since the predictions systematically underestimate
the exact values.

\subsection{Construction of the semi-local functional at finite temperature}
\begin{figure}
	\centering
	\includegraphics[width=\columnwidth]{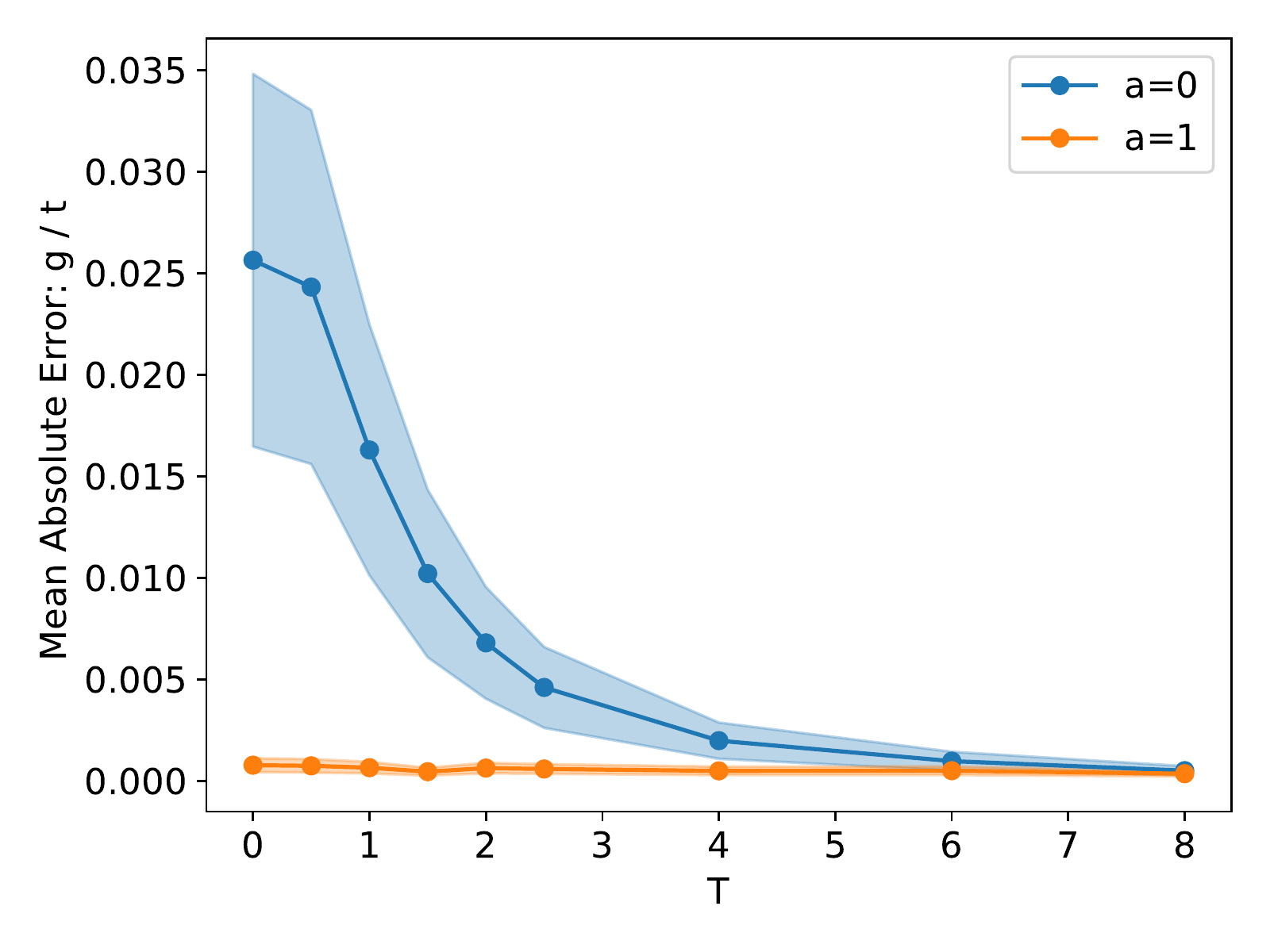}
	\caption{(color on-line) MAE of the finite-temperature universal functional, $g_V$, as a function of $T$ and for different
	value of the locality parameter, $a$. Data are presented for $L=6$ rings and parameters $V=1$ and $W=4$. The equilibrium site 
	occupation used to evaluate the functional is that obtained by exact diagonalization. The shaded regions are the standard deviation 
	of the absolute error. For each temperature the training, validation and test sets contain, 6,000, 2,000 and 2,000 samples, 
	respectively.}
	\label{fig:fig6_local}
\end{figure}

Finally, we discuss the extension of ML lattice DFT to finite temperature, namely we compute the universal functional, $g_V$.
Firstly, we investigate the locality of the functional by evaluating the MAE as a function of temperature for $a=0$ and $a=1$, 
when $g_V$ is computed at the exact equilibrium site-occupation, $\{n_i^T\}$. Our results for $L=6$, $V=1$ and $W=4$ are 
presented in figure~\ref{fig:fig6_local}. For each temperature the model has been trained over 6,000 samples, validated over 
2,000 and tested over 2,000. Two considerations can be made. On the one hand, a semi-local approximation clearly
performs better than a fully local one, which underperforms, in particular at low temperature. On the other hand, the error is
significantly reduced as the temperature gets larger and already at $T>6$ little difference is detected between a semi-local and 
a local approximation. In other words, the functional becomes more local as the temperature increases. This should not be 
surprising considering that in the limit $T\rightarrow\infty$ the equilibrium density matrix becomes 
\begin{equation}
	\hat{\rho} = \frac{1}{\Omega} \mathbb{1}\:,
\end{equation}
where $\Omega$ is the dimension of the Hilbert space and $\mathbb{1}$ the identity. This means that for $T\rightarrow\infty$ the
equilibrium site occupation is uniform, $n_i^T=N_e/L$, and hence a local and semi-local functionals contain the same information.

\begin{figure}
	\centering
	\includegraphics[width=\columnwidth]{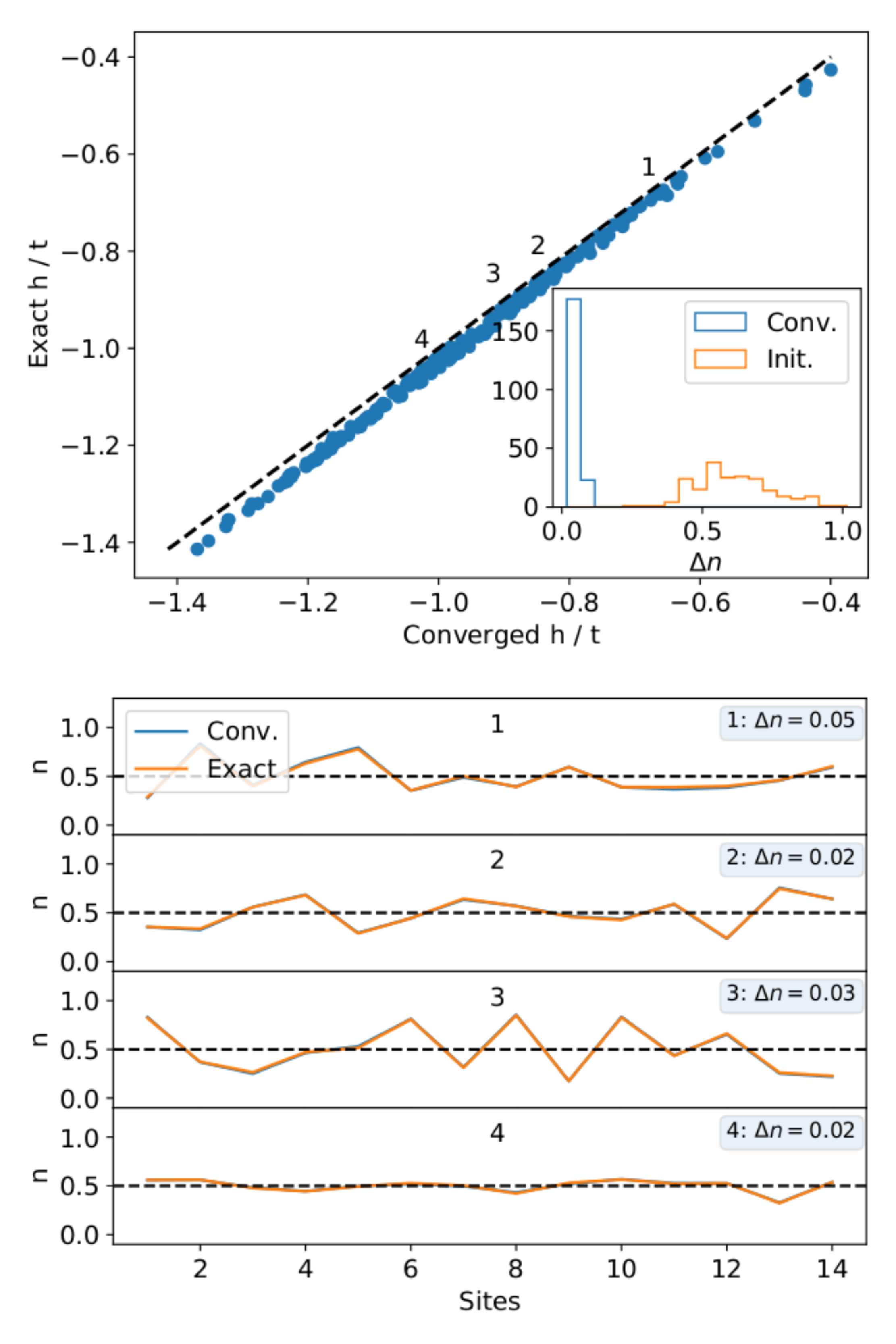}
	\caption{(color on-line) Demonstration of the variational principle for the finite-temperature ML functional. In the top panel we show the 
	Helmholtz free energy density computed by minimization with respect to the site occupation. Our network has been trained for $L=10$ 
	and used for a ring containing 14 sites ($V=1$). The figure contains 100 different disorder realizations with $W=4$. 
	The insert shows histograms of the euclidean distances $\Delta n$, of the initial and converged equilibrium site occupation from the 
	exact one. The lower panel presents a selection of the converged occupations compared against the exact value. These are taken for 
	different ground-state energy densities as indicated by the numbers in the upper panel.
	}
	\label{fig:fig7_gdFT}
\end{figure}

Finally, we demonstrate the variational principle for finite temperature DFT with a semi-local functional. Also in this case we train on $L=10$ 
systems and compute the equilibrium site occupation for rings with 14 sites by gradient-descent minimization (we use exactly the same
algorithm as in the $T=0$ case). The starting site occupation is uniform, and the converged one is finally used to compute the 
Helmholtz free energy density. Figure~\ref{fig:fig7_gdFT} presents our results for $V=1$, $W=4$, $a=2$ and temperatures randomly distributed 
between 1 and 2. 

Clearly, the ML finite-temperature functional appears able to return equilibrium site occupations extremely close to the exact ones
(the MAE computed over 100 configurations is 0.029), and good quality Helmholtz free energy densities. Intriguingly, the euclidian 
distance between the converged site occupations and the exact ones is much smaller than in the $T=0$ limit (compare the histograms 
in Fig.~\ref{fig:fig4_gd} and Fig.~\ref{fig:fig7_gdFT}). This reflects two main facts. Firstly, the finite-temperature site occupation is, on 
average, closer to the homogeneous case than the zero-$T$ one; secondly the functional becomes more local at finite temperature. 
Such tendency towards relatively homogeneous site occupations at high temperatures brings back the issue related to the incorrect 
homogeneous limit of the semi-local funcitonals, discussed for the $T=0$ case. In fact, the Helmholtz free energy density appears 
now systematically underestimated, in particular for rather uniform densities (see lower panel 4 of Fig.~\ref{fig:fig7_gdFT}). Again the 
issue can be resolved by either training on larger rings, or by subtracting the non-interacting Helmholtz free energy density from the 
functional.

\section{Conclusion and Future Work}

In this work we have introduced a semi-local approximation to lattice density functional theory for the spinless Hubbard model in one
dimension. This has been defined for ground-state DFT, and it has been extended to finite temperature, where the universal functional
is constructed for the Helmholtz free energy density. At the same time, strong of the Hohenberg-Kohn theorems, we have evaluated a 
number of one-to-one mappings between the ground-state site occupation and various thermodynamical quantities at finite temperature.
Also in this case a semi-local mapping appears to be possible. 

The benefit of constructing such semi-local approximations is that one can train the machine-learning models over systems of moderate 
size and then use them for much larger ones. In practice, this allows one to explore the thermodynamical limit of various interacting 
models with disorder. It must be noted that the semi-local approximation defined here does not describe the homogeneous limit, a problem 
that becomes evident when training on very small systems, or for rather homogeneous exact density. Such problem can be overcome 
by an appropriate subtraction of the non-interacting energy density (or of the non-interacting Helmholtz free energy density), or simply 
by training over systems with adequate size. 

The work presented here can extend now to several future directions. Firstly, one can construct similar formalism in two and three dimensions.
In this case the dimension of Hilbert space becomes prohibitively large already for small systems and it is likely that it may be no longer possible 
to generate the training set by exact diagonalization. However, one can alternatively use datasets generated with many-body techniques
such as Quantum Montecarlo or other real space approaches (e.g. the Hubbard-Stratonovich approach). A second possibility is to incorporate 
spin into the representation, namely consider spin-full models (e.g. the complete Hubbard model). All in all, it appears that machine-learning 
lattice DFT may become a powerful tool to explore the interplay between interaction and disorder in different dimensions and various lattice 
models.

\subsubsection*{Acknowledgements}
Funding is provided by the Irish Research Council (JN) and by the European Research Council project {\sc quest} (RT).
We acknowledge the DJEI/DES/SFI/HEA Irish Centre for High-End Computing (ICHEC) and Trinity Centre for High 
Performance Computing (TCHPC) for  the provision of computational resources. Special thanks to Dr. Alessandro Lunghi for 
reading an early draft of the paper.

\bibliography{paper}

\end{document}